# On the remarkable properties of Weyl particles

Georgios N. Tsigaridas[1,*], Aristides I. Kechriniotis[2], Christos A. Tsonos[2] and Konstantinos K. Delibasis[3]

[1]Department of Physics, School of Applied Mathematical and Physical Sciences, National Technical University of Athens, GR-15780 Zografou Athens, Greece

[2]Department of Physics, University of Thessaly, GR-35100 Lamia, Greece

[3]Department of Computer Science and Biomedical Informatics, University of Thessaly, GR-35131 Lamia, Greece

[*]Corresponding Author. E-mail: gtsig@mail.ntua.gr

**Abstract**

In this work we show that Weyl particles can exist at different states in zero electromagnetic field, either as free particles, or at localized states described by a parameter with dimensions of mass. We also calculate the electromagnetic fields that should be applied in order to modify the localization of Weyl particles at a desired rate. It is shown that they are simple electric fields, which can be easily implemented experimentally. Consequently, the localization of Weyl particles in certain materials supporting these particles could also be studied experimentally, in the framework of solid-state physics or in the framework of laser physics, using ions trapped by laser beams. In addition, a particularly important remark is that the localization of the energy of the particles can lead to the generation of gravitational mass, according to Einstein's field equations of general relativity. Furthermore, in the case that the energy and localization of the particles exceeds a critical level, tiny black holes could also be created, potential candidates for the dark matter of the universe.

1. Introduction

In our previous work [1] we had shown that all spinors of the form

$$\psi = \begin{pmatrix} \cos\left(\dfrac{\theta(t)}{2}\right) \\ e^{i\varphi(t)} \sin\left(\dfrac{\theta(t)}{2}\right) \end{pmatrix} \exp\left[ih(\mathbf{r},t)\right], \qquad (1)$$

where $\theta(t), \varphi(t)$ are arbitrary real functions of time and $h(\mathbf{r},t) = h(x,y,z,t)$ is an arbitrary real function of the spatial coordinates and time are solutions to the Weyl equation

$$i\sigma^\mu \partial_\mu \psi + a_\mu \sigma^\mu \psi = 0 \tag{2}$$

for the 4-potentials

$$(a_0, a_1, a_2, a_3) = \left( \frac{\partial h}{\partial t} + \frac{1}{2}\frac{d\varphi}{dt}, \frac{\partial h}{\partial x} + \frac{1}{2}\sin\varphi \frac{d\theta}{dt}, \frac{\partial h}{\partial y} - \frac{1}{2}\cos\varphi \frac{d\theta}{dt}, \frac{\partial h}{\partial z} - \frac{1}{2}\frac{d\varphi}{dt} \right) \tag{3}$$

Here, $\sigma^\mu$ are the standard Pauli matrices [2] and $a_\mu = qA_\mu$, where $q$ is the electric charge of the particles and $A_\mu$ is the electromagnetic 4-potential. Eq. (2) is expressed in natural units, where $\hbar = c = 1$, and describes particles with positive helicity.

In addition, according to theorem 3.1 in [3], the spinors (1) will also be solutions to the Weyl equation (2) for an infinite number of 4-potentials given by the formula

$$b_\mu = a_\mu + \kappa_\mu s, \ \mu = 0, 1, 2, 3 \tag{4}$$

where

$$\begin{aligned}(\kappa_0, \kappa_1, \kappa_2, \kappa_3) &= \left( 1, -\frac{\psi^\dagger \sigma^1 \psi}{\psi^\dagger \psi}, -\frac{\psi^\dagger \sigma^2 \psi}{\psi^\dagger \psi}, -\frac{\psi^\dagger \sigma^3 \psi}{\psi^\dagger \psi} \right) \\ &= (1, -\sin\theta\cos\varphi, -\sin\theta\sin\varphi, -\cos\theta) \end{aligned} \tag{5}$$

and $s$ is an arbitrary real function of the spatial coordinates and time.

Similarly, all spinors of the form

$$\psi' = \begin{pmatrix} -\sin\left(\frac{\theta(t)}{2}\right) \\ e^{i\varphi(t)} \cos\left(\frac{\theta(t)}{2}\right) \end{pmatrix} \exp[ih(\mathbf{r}, t)] \tag{6}$$

are solutions to the Weyl equation

$$i\sigma^\mu \partial_\mu \psi - 2i\sigma^0 \partial_0 \psi + a_\mu \sigma^\mu \psi - 2a_0 \sigma^0 \psi = 0 \tag{7}$$

for the 4-potentials

$$(a'_0, a'_1, a'_2, a'_3) = \left( \frac{\partial h}{\partial t} + \frac{1}{2}\frac{d\varphi}{dt}, \frac{\partial h}{\partial x} - \frac{1}{2}\sin\varphi \frac{d\theta}{dt}, \frac{\partial h}{\partial y} + \frac{1}{2}\cos\varphi \frac{d\theta}{dt}, \frac{\partial h}{\partial z} + \frac{1}{2}\frac{d\varphi}{dt} \right) \tag{8}$$

The Weyl equation in the form of (7) corresponds to particles with negative helicity and it can also be written as

$$i\sigma'^\mu \partial_\mu \psi + a_\mu \sigma'^\mu \psi = 0 \tag{9}$$

where $\sigma'^0 = \sigma^0, \sigma'^1 = -\sigma^1, \sigma'^2 = -\sigma^2, \sigma'^3 = -\sigma^3$.

The spinors of (6) will also satisfy the Weyl equation (7), or equivalently (9), for an infinite number of 4-potentials, given by the formula

$$b'_\mu = a'_\mu + \kappa'_\mu s(\mathbf{r},t), \quad \mu = 0,1,2,3 \tag{10}$$

where

$$(\kappa'_0, \kappa'_1, \kappa'_2, \kappa'_3) = (1, -\sin\theta\cos\varphi, -\sin\theta\sin\varphi, -\cos\theta) = (\kappa_0, \kappa_1, \kappa_2, \kappa_3) \tag{11}$$

Assuming that $\dfrac{d\theta}{dt} = \dfrac{d\varphi}{dt} = 0$ and setting

$$h(\mathbf{r},t) = E_0 \left[ x\sin\theta\cos\varphi + y\sin\theta\sin\varphi + z\cos\theta - t \right] \tag{12}$$

the above solutions correspond to free Weyl particles with energy $E_0$ moving along a straight line in space with polar angle $\theta$ and azimuthal angle $\varphi$. However, these solutions are much more general and have very important and unexpected properties, as it will be shown in the following sections.

## 2. On the property of Weyl particles to exist at different quantum states in zero electromagnetic field

The electromagnetic fields corresponding to the 4-potentials (3), (8) can be easily calculated through the formulae [4]

$$\mathbf{E} = -\nabla U - \frac{\partial \mathbf{A}}{\partial t}, \quad \mathbf{B} = \nabla \times \mathbf{A} \tag{13}$$

where $U = b_0/q$ is the electric potential and $\mathbf{A} = -(1/q)(b_1\mathbf{i} + b_2\mathbf{j} + b_3\mathbf{k})$ is the magnetic vector potential. The choice of the minus sign in the definition of the magnetic potential is related to the form of the Weyl equations used in this article.

Thus, in the case of particles with positive helicity, the electromagnetic field corresponding to the 4-potential $a_\mu$ is

$$\begin{aligned}
\mathbf{E} &= \frac{1}{2q}\left(\cos\varphi\frac{d\theta}{dt}\frac{d\varphi}{dt} + \sin\varphi\frac{d^2\theta}{dt^2}\right)\mathbf{i} \\
&+ \frac{1}{2q}\left(\sin\varphi\frac{d\theta}{dt}\frac{d\varphi}{dt} - \cos\varphi\frac{d^2\theta}{dt^2}\right)\mathbf{j} \\
&- \frac{1}{2q}\frac{d^2\varphi}{dt^2}\mathbf{k} \\
\mathbf{B} &= \mathbf{0}
\end{aligned} \tag{14}$$

Similarly, in the case of particles with negative helicity, the electromagnetic field corresponding to the 4-potential $a'_\mu$ is $\mathbf{E}' = -\mathbf{E}$ and $\mathbf{B}' = \mathbf{0}$.

Obviously, the above electromagnetic fields become zero in the case that

$$\frac{d^2\theta}{dt^2} = \frac{d^2\varphi}{dt^2} = \frac{d\theta}{dt}\frac{d\varphi}{dt} = 0. \tag{15}$$

implying that Weyl particles, in zero electromagnetic field, can either move on a straight line $(d\theta/dt = d\varphi/dt = 0)$, or move with constant angular velocity regarding the polar angle $(d\theta/dt = \omega_1, \ d\varphi/dt = 0)$, or move with constant angular velocity with respect to the azimuthal angle $(d\theta/dt = 0, \ d\varphi/dt = \omega_2)$.

In more detail, the velocity of the particles, can be defined as in [5, 6] in the case of positive helicity

$$\begin{aligned}\mathbf{v} &= \left(\psi^\dagger \sigma^0 \sigma^1 \psi\right)\mathbf{i} + \left(\psi^\dagger \sigma^0 \sigma^2 \psi\right)\mathbf{j} + \left(\psi^\dagger \sigma^0 \sigma^3 \psi\right)\mathbf{k} \\ &= \sin\theta\cos\varphi\,\mathbf{i} + \sin\theta\sin\varphi\,\mathbf{j} + \cos\theta\,\mathbf{k}\end{aligned} \tag{16}$$

and as following in the case of negative helicity

$$\begin{aligned}\mathbf{v}' &= \left(\psi'^\dagger \sigma'^0 \sigma'^1 \psi'\right)\mathbf{i} + \left(\psi'^\dagger \sigma'^0 \sigma'^2 \psi'\right)\mathbf{j} + \left(\psi'^\dagger \sigma'^0 \sigma'^3 \psi'\right)\mathbf{k} \\ &= \sin\theta\cos\varphi\,\mathbf{i} + \sin\theta\sin\varphi\,\mathbf{j} + \cos\theta\,\mathbf{k} \\ &= \mathbf{v}\end{aligned} \tag{17}$$

Therefore, in zero electromagnetic field, it can either become constant assuming that $d\theta/dt = d\varphi/dt = 0$, or take the form

$$\mathbf{v}_\theta = \sin(\theta_0 + \omega_1 t)\cos\varphi_0 \mathbf{i} + \sin(\theta_0 + \omega_1 t)\sin\varphi_0 \mathbf{j} + \cos(\theta_0 + \omega_1 t)\mathbf{k} \tag{18}$$

in the case that $d\theta/dt = \omega_1, \ d\varphi/dt = 0$, or the form

$$\mathbf{v}_\varphi = \sin\theta_0 \cos(\varphi_0 + \omega_2 t)\mathbf{i} + \sin\theta_0 \sin(\varphi_0 + \omega_2 t)\mathbf{j} + \cos\theta_0 \mathbf{k} \tag{19}$$

in the case that $d\theta/dt = 0, \ d\varphi/dt = \omega_2$. In the above expressions $\theta_0$, $\varphi_0$ are arbitrary reals constants, corresponding to the polar and azimuthal angles of the particles respectively, in the case that they do not change with time.

It is easy to verify that a classical particle with velocity $\mathbf{v}_\theta$ performs circular motion with radius $\omega_1^{-1}$, while a classical particle with velocity $\mathbf{v}_\varphi$ performs helicoidal motion with radius $\omega_2^{-1}\sin\theta_0$ and pitch $\pi\omega_2^{-1}\sin 2\theta_0$. Obviously, in the case that $\theta_0 = \pi/2$ the particle still performs circular motion with radius $\omega_2^{-1}$. Consequently, Weyl particles in zero electromagnetic field, can either behave as free particles assuming that $d\theta/dt = d\varphi/dt = 0$, or exist at a localized bound state in the case that $d\theta/dt = \omega_1, \ d\varphi/dt = 0$, or exist at an intermediate state, bound on the x-y plane, and free along the z-axis, corresponding to $d\theta/dt = 0, \ d\varphi/dt = \omega_2$. Thus, Weyl particles have the remarkable property to exist at different states, not equivalent to one another, in the same zero electromagnetic field.

It is important to note that, in all cases, the modulus of the velocity of the particles is equal to one in natural units, as required by the special theory of relativity. Finally, it should be mentioned that the state of the particles will not be affected in the case that the following electromagnetic fields

$$\mathbf{E}_s(\mathbf{r},t) = -\frac{1}{q}\left[\sin\theta\cos\varphi\frac{\partial s}{\partial t} + \frac{\partial s}{\partial x} + s\left(\cos\theta\cos\varphi\frac{d\theta}{dt} - \sin\theta\sin\varphi\frac{d\varphi}{dt}\right)\right]\mathbf{i}$$
$$-\frac{1}{q}\left[\sin\theta\sin\varphi\frac{\partial s}{\partial t} + \frac{\partial s}{\partial y} + s\left(\cos\theta\sin\varphi\frac{d\theta}{dt} + \sin\theta\cos\varphi\frac{d\varphi}{dt}\right)\right]\mathbf{j}$$
$$-\frac{1}{q}\left(\cos\theta\frac{\partial s}{\partial t} + \frac{\partial s}{\partial z} + \sin\theta\frac{d\theta}{dt}s\right)\mathbf{k} \tag{20}$$

$$\mathbf{B}_s(\mathbf{r},t) = \frac{1}{q}\left(-\sin\theta\sin\varphi\frac{\partial s}{\partial z} + \cos\theta\frac{\partial s}{\partial y}\right)\mathbf{i} + \frac{1}{q}\left(\sin\theta\cos\varphi\frac{\partial s}{\partial z} - \cos\theta\frac{\partial s}{\partial x}\right)\mathbf{j}$$
$$+\frac{1}{q}\sin\theta\left(-\cos\varphi\frac{\partial s}{\partial y} + \sin\varphi\frac{\partial s}{\partial x}\right)\mathbf{k}$$

corresponding to the 4-potentials

$$(U, \mathbf{A}) = (1, \sin\theta\cos\varphi, \sin\theta\sin\varphi, \cos\theta)s(\mathbf{r},t)/q \tag{21}$$

are added to **E**, **B** or **E**′, **B**′.

It is important to note that, according to the above equation, the magnetic vector potential is the product of the electric scalar potential with the velocity of the particles. This also happens in the case of the Liénard–Wiechert potentials [7], describing the classical electromagnetic effect of a moving electric point charge in terms of the magnetic vector potential and the electric scalar potential in the Lorenz gauge. In natural units, the magnetic potential is connected to the electric potential through the formula $\mathbf{A}(\mathbf{r},t) = \mathbf{v}_s(t_r)U(\mathbf{r},t)$, where $\mathbf{v}_s(t_r)$ is the velocity of the source charge, evaluated at the retarded time, $t_r = t - |\mathbf{r} - \mathbf{r}_s(t_r)|$. Here, **r** is the position of the observation point and $\mathbf{r}_s$ is the position of the source charge point, at the time of the signal's origin from that point.

Thus, assuming that we have two localized Weyl particles, moving in closed circular orbits, in order the 4-potential (21) to be equal to the Liénard–Wiechert potential, the distance between the two particles, defined as the distance between the centers of the two circular orbits, should be equal to integer multiples of $2\pi r$, where $r$ is the radius of each circular orbit. Strictly speaking, the term orbit refers to the trajectory of a classical particle with the same velocity as that of the Weyl particle. Thus, if Weyl particles are arranged on a lattice of step equal to integer multiples of $2\pi r$, then the interaction between the particles will not have any influence on their quantum state.

This is another interesting implication of the fact that Weyl spinors are degenerate, as shown in theorem 3.1 in [3].

In the rest of the paper we shall assume that the interaction between Weyl particles is negligible. This assumption is justified by the fact that Weyl particles are elementary and consequently their charge is expected to be of the order of the electron charge. Therefore, the potentials and fields generated by these particles are expected to be negligible compared to the external fields. We shall also assume that the function $s$ depends only on time, in order to simplify the calculations and make the physical interpretation of the results more transparent.

In conclusion, in this section we have shown that Weyl particles can be considered as double degenerate, in the sense that they can exist at the same state in a wide variety of electromagnetic fields, and at the same time, they can exist at different states in zero electromagnetic field. Although this property alone is quite remarkable, in the following sections we shall show that Weyl particles have additional particularly interesting properties, related to their localization and mass generation.

### 3. On the localization of Weyl particles

The kinetic 4-momentum corresponding to the spinors in (1) and (6) is

$$\pi_t = \psi^\dagger \left( -i \frac{\partial}{\partial t} - b_0 \right) \psi = -\frac{1}{2} \cos\theta \frac{d\varphi}{dt} - s$$

$$\pi_x = \psi^\dagger \left( -i \frac{\partial}{\partial x} - b_1 \right) \psi = -\frac{1}{2} \sin\varphi \frac{d\theta}{dt} + s \sin\theta \cos\varphi$$

$$\pi_y = \psi^\dagger \left( -i \frac{\partial}{\partial y} - b_2 \right) \psi = \frac{1}{2} \cos\varphi \frac{d\theta}{dt} + s \sin\theta \sin\varphi \qquad (22)$$

$$\pi_z = \psi^\dagger \left( -i \frac{\partial}{\partial z} - b_3 \right) \psi = \frac{1}{2} \frac{d\varphi}{dt} + s \cos\theta$$

and

$$\pi'_t = \psi'^\dagger \left( -i \frac{\partial}{\partial t} - b'_0 \right) \psi' = \frac{1}{2} \cos\theta \frac{d\varphi}{dt} - s$$

$$\pi'_x = \psi'^\dagger \left( -i \frac{\partial}{\partial x} - b'_1 \right) \psi' = \frac{1}{2} \sin\varphi \frac{d\theta}{dt} + s \sin\theta \cos\varphi$$

$$\pi'_y = \psi'^\dagger \left( -i \frac{\partial}{\partial y} - b'_2 \right) \psi' = -\frac{1}{2} \cos\varphi \frac{d\theta}{dt} + s \sin\theta \sin\varphi \qquad (23)$$

$$\pi'_z = \psi'^\dagger \left( -i \frac{\partial}{\partial z} - b'_3 \right) \psi' = -\frac{1}{2} \frac{d\varphi}{dt} + s \cos\theta$$

respectively [5]. Thus, the energy and momentum of the particles in the case of positive and negative helicity take the following forms

$$E_0 = \pi_t = -\frac{1}{2}\cos\theta\frac{d\varphi}{dt} - s$$

$$\mathbf{p} = -(\pi_x\mathbf{i} + \pi_y\mathbf{j} + \pi_z\mathbf{k}) = -\left(-\frac{1}{2}\sin\varphi\frac{d\theta}{dt} + s\sin\theta\cos\varphi\right)\mathbf{i} \quad (24)$$

$$-\left(\frac{1}{2}\cos\varphi\frac{d\theta}{dt} + s\sin\theta\sin\varphi\right)\mathbf{j} - \left(\frac{1}{2}\frac{d\varphi}{dt} + s\cos\theta\right)\mathbf{k}$$

and

$$E'_0 = \pi'_t = \frac{1}{2}\cos\theta\frac{d\varphi}{dt} - s$$

$$\mathbf{p}' = -(\pi'_x\mathbf{i} + \pi'_y\mathbf{j} + \pi'_z\mathbf{k}) = -\left(\frac{1}{2}\sin\varphi\frac{d\theta}{dt} + s\sin\theta\cos\varphi\right)\mathbf{i} \quad (25)$$

$$-\left(-\frac{1}{2}\cos\varphi\frac{d\theta}{dt} + s\sin\theta\sin\varphi\right)\mathbf{j} - \left(-\frac{1}{2}\frac{d\varphi}{dt} + s\cos\theta\right)\mathbf{k}$$

From the above expressions it is clear that it is possible to fully control the energy of the particles through the function $s(t)$, applying the electromagnetic field of (20). Specifically, the rate of change of the energy of the particles becomes

$$\frac{dE_0}{dt} = \frac{1}{2}\sin\theta\frac{d\theta}{dt}\frac{d\varphi}{dt} - \frac{1}{2}\cos\theta\frac{d^2\varphi}{dt^2} - \frac{ds}{dt} \quad (26)$$

and

$$\frac{dE'_0}{dt} = -\frac{1}{2}\sin\theta\frac{d\theta}{dt}\frac{d\varphi}{dt} + \frac{1}{2}\cos\theta\frac{d^2\varphi}{dt^2} - \frac{ds}{dt} \quad (27)$$

for positive and negative helicity respectively. It is easy to confirm that if condition (15) is satisfied, the rate of the energy change becomes $dE/dt = -ds/dt$. In this case the electromagnetic fields in (20) take the simple form

$$\mathbf{E}_s(t) = \frac{1}{q}\frac{dE}{dt}(\sin\theta\cos\varphi\mathbf{i} + \sin\theta\sin\varphi\mathbf{j} + \cos\theta\mathbf{k}) \quad (28)$$

$$\mathbf{B}_s(t) = \mathbf{0}$$

Thus, the energy of the Weyl particles can be fully controlled by applying an electric field along the propagation direction of the particles, depending on the direction of the field relative to the propagation direction of the particles. In addition, assuming that the electric field is constant, Eq. (28) implies that the rate of the energy change is also constant, equal to $q|\mathbf{E}|$, where $|\mathbf{E}|$ is the magnitude of the electric field. Obviously, the energy increases if the field is applied parallel to the propagation direction of the particles and decreases if applied in the opposite direction. In S.I. units the rate of the energy change in $J/s$ is equal to $q|\mathbf{E}|c$, where $|\mathbf{E}|$ is the magnitude of the electric field in $V/m$, $q$ is the charge of the particle in $C$ and $c$ the speed of light in $m/s$. If we further assume that the charge of the particles is equal to the electron charge, the above analysis implies that the energy of the particles changes at a rate of $c|\mathbf{E}|$ $eV/s$, or $|\mathbf{E}|$ $eV$ per meter of propagation inside the constant field. Thus, applying a constant electric field for the appropriate amount of time, it is even possible to annihilate the particle by making its energy zero. If the field continues to be applied, the particle will reappear moving in the opposite direction and gaining energy with time.

The above analysis is also valid for massless Dirac particles described by degenerate spinors. However, in this case, it is not possible to change the propagation direction pf the particles without changing the form of the solutions.

As far as the energy and momentum of Weyl particles is concerned, it is easy to verify that they obey the basic laws of physics, namely

$$\frac{d\mathbf{p}}{dt} = \mathbf{F} = q\mathbf{E}, \quad \frac{d\mathbf{p}'}{dt} = \mathbf{F}' = q\mathbf{E}' \tag{29}$$

and

$$\frac{dE_0}{dt} = \mathbf{F} \cdot \mathbf{v} = q\mathbf{E} \cdot \mathbf{v}, \quad \frac{dE_0'}{dt} = \mathbf{F}' \cdot \mathbf{v}' = q\mathbf{E}' \cdot \mathbf{v}' \tag{30}$$

where $\mathbf{F}, \mathbf{F}'$ is the electromagnetic force exerted to particles with positive and negative helicity respectively.

However, in both cases, the energy of the particles is smaller than the modulus of their momentum, except in the special case the particles move on straight lines $(d\theta/dt = d\varphi/dt = 0)$. Specifically

$$E_0^2 - |\mathbf{p}|^2 = E_0'^2 - |\mathbf{p}'|^2 = -\frac{1}{4}\left[\sin^2\theta\left(\frac{d\varphi}{dt}\right)^2 + \left(\frac{d\theta}{dt}\right)^2\right] \tag{31}$$

This result indicates that, as the particles change their propagation direction, they behave as having imaginary mass given by the formula

$$m^* = \frac{i}{2}\left[\sin^2\theta\left(\frac{d\varphi}{dt}\right)^2 + \left(\frac{d\theta}{dt}\right)^2\right]^{1/2} \tag{32}$$

Thus, we can define a parameter $k$ as

$$k = -i\,m^* = \frac{1}{2}\left[\sin^2\theta\left(\frac{d\varphi}{dt}\right)^2 + \left(\frac{d\theta}{dt}\right)^2\right]^{1/2} \tag{33}$$

which has dimensions of mass and is a measure of the rate of change of the propagation direction of the particles.

Here, it should be noted that particles in zero electromagnetic field can have several values of parameter $k$, namely $k=0$ in the case that $d\theta/dt = d\varphi/dt = 0$, or $k=|\omega_1|/2$ in the case that $d\theta/dt = \omega_1$, $d\varphi/dt = 0$, or $k=|\omega_2 \sin\theta_0|/2$ in the case that $d\theta/dt = 0$, $d\varphi/dt = \omega_2$. Also, since the radius of the circle or the helix, corresponding to the cases that $d\theta/dt = \omega_1$, $d\varphi/dt = 0$ and $d\theta/dt = 0$, $d\varphi/dt = \omega_2$ respectively, is inverse proportional to the angular velocity, we conclude that the values of the parameter $k$ increase as the particles become more localized. Therefore, although $k$ has dimensions of mass, it is a measure of the localization of the particles, and it is not directly associated with real gravitational mass. Consequently, it will be mentioned as localization parameter in the following. More details on the connection between the localization of the particles and the generation of real gravitational mass will be provided in the next section.

Another important remark is that the vector of the momentum, as defined by equations (24), (25) is not collinear to the vector of the velocity, as defined by equations (18), (19). For example, supposing that $d\theta/dt = 0$, $\theta_0 = \pi/2$, the velocity of the particle becomes $\mathbf{v} = \cos\varphi\mathbf{i} + \sin\varphi\mathbf{j}$, while the momentum takes the form $\mathbf{p} = -s\cos\varphi\mathbf{i} - s\sin\varphi\mathbf{j} - (1/2)(d\varphi/dt)\mathbf{k}$, or $\mathbf{p} = E_0\cos\varphi\mathbf{i} + E_0\sin\varphi\mathbf{j} - (1/2)(d\varphi/dt)\mathbf{k}$ where $E_0 = -s$ is the energy of the particle. Consequently, although the velocity of the particle lies on the x-y plane, the z-component of the momentum is non-zero. Also, an interesting remark is that the quantity $|\mathbf{p}\times\mathbf{v}|^2$ is equal to $|\mathbf{p}|^2 - E_0^2 = k^2$. Thus, the higher the deviation of the particle from the linear motion, the higher the non-collinearity between the velocity and the momentum of the particle becomes.

This effect, as well as the discrepancy between the modulus of the energy and the momentum described by Eq. (31) can be attributed to fact that the energy and the momentum are not well defined for massless particles. In more detail, according to the special theory of relativity, the energy and momentum of a massive particle are given, in S.I. units, by the formulae $E = \gamma mc^2$ and $\mathbf{p} = \gamma m\mathbf{v}$, respectively. Here $m$ is the mass of the particle and $\gamma = (1-|\mathbf{v}|^2/c^2)^{-1}$. Thus, in the case of massless particles $(m=0)$ both the energy and the momentum are not well defined, taking the indeterminate form $0/0$. In addition, in the case that the propagation direction of the particle changes, the reference frame of the particle is not inertial regarding the reference frame of the laboratory. Consequently, the laws of physics are expected to take different forms in the two frames of reference and Weyl equation is expected to be valid locally. Indeed, it can be considered that the particle in its reference frame moves momentarily as free particle with energy $E_0$ and momentum $\mathbf{p} = E_0 \sin\theta\cos\varphi\mathbf{i} + E_0\sin\theta\sin\varphi\mathbf{j} + E_0\cos\theta\mathbf{k}$, collinear to its velocity.

Furthermore, the fact that the difference $E_0^2 - |\mathbf{p}|^2$ is negative may seem peculiar at first glance, but it could be interpreted as a result of the localization of the particles, which increases the uncertainty in their momentum. Indeed, supposing that $d\theta/dt = \omega_1$, $d\varphi/dt = 0$, Eq. (31) becomes

$$E_0^2 - |\mathbf{p}|^2 = -\frac{1}{4}\omega_1^2 = -\frac{1}{d^2} \tag{34}$$

where $d$ is a measure of the localization of the particle, strictly defined as the diameter of the trajectory of a classical particle moving with the same velocity as the Weyl one. Thus, the difference $E_0^2 - |\mathbf{p}|^2$ is inversely proportional to the localization of the particles. However, at the same time, due to the Heisenberg's uncertainty principle, the uncertainty in the momentum of the particles also increases. In more detail, setting $|\mathbf{p}|^2 = (p_0 + \Delta p)^2$ with $p_0 = E_0$ in Eq. (31) yields

$$2p_0 d(d\Delta p) + (d\Delta p)^2 = 1 \tag{35}$$

which can be easily solved for $(d\Delta p)$. The positive solution is

$$d\Delta p = -p_0 d + \sqrt{1 + (p_0 d)^2} \tag{36}$$

Thus, the quantity $d\Delta p$ takes values in the range from zero, as $p_0 d \to \infty$, to one as $p_0 d \to 0$. However, as mentioned above, the parameter $d$ is a measure of the uncertainty in the position of the particle, and consequently, according to Heisenberg's uncertainty principle, the product $d\Delta p$ should be of the order of $\hbar/2$, or $1/2$ in natural units, in agreement to what is predicted by Eq. (36). A similar analysis can be followed in the case that $d\theta/dt = 0$, $d\varphi/dt = \omega_2$, or in the more general case that $d\theta/dt = \omega_1$, $d\varphi/dt = \omega_2$. The above analysis indicates that the localized states could be stable, allowed by the position – momentum uncertainty principle.

To gain more insight on the behavior of Weyl particles in localized states, we consider that the angles $\theta$, $\varphi$ change at a constant rate, namely $d\theta/dt = \omega_1$, $d\varphi/dt = \omega_2$. In this case, the parameter $k$ becomes

$$k = \frac{1}{2}\left[\omega_1^2 + \omega_2^2 \sin^2(\theta_0 + \omega_1 t)\right]^{1/2} \tag{37}$$

oscillating between the values $|\omega_1|/2$ and $\left(\omega_1^2 + \omega_2^2\right)^{1/2}/2$.

As an example, in figure 1 we provide a parametric plot of the velocity of the particle, as given by Eq. (16), in the time interval $t \in [0, 200]$, supposing that $\omega_1 = \sqrt{3}$, $\omega_2 = \sqrt{5}$, $\varphi_0 = 0$ and $\theta_0 = \pi/2$.

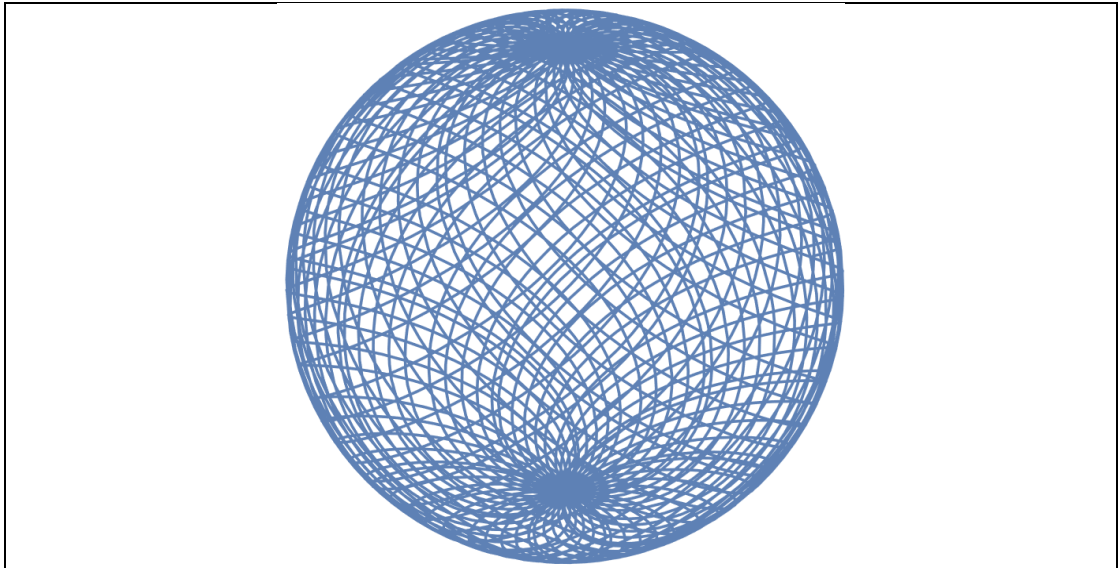

**Figure 1:** Parametric plot of the velocity of a Weyl particle in the time interval $t \in [0, 200]$, in the case that $\omega_1 = \sqrt{3}$, $\omega_2 = \sqrt{5}$, $\varphi_0 = 0$ and $\theta_0 = \pi/2$.

It is clear that the motion of the Weyl particle is exceptionally complex, as it is also shown in figure 2, depicting the trajectory of a classical particle with the same velocity as that of the Weyl particle. However, at the same time, it is evident that the motion is bound and consequently, the Weyl particle is localized, with the values of the localization parameter $k$ oscillating in the interval $k \in \left[\sqrt{3}/2, \sqrt{2}\right]$ as shown in figure 3.

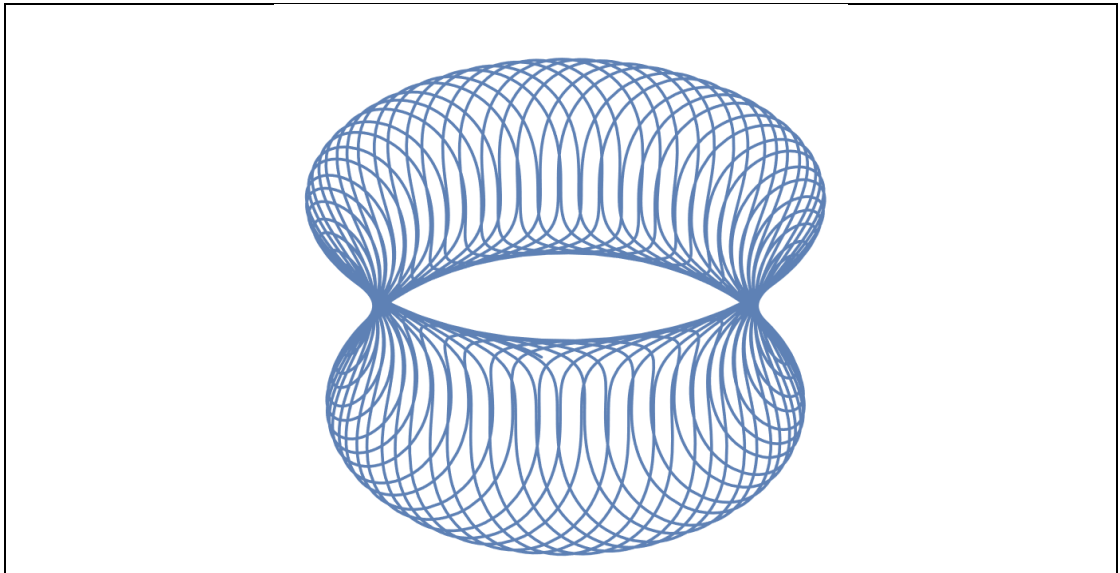

**Figure 2:** The trajectory of a classical particle with the same velocity as the Weyl particle in the time interval $t \in [0, 200]$, in the case that $\omega_1 = \sqrt{3}$, $\omega_2 = \sqrt{5}$, $\varphi_0 = 0$ and $\theta_0 = \pi/2$.

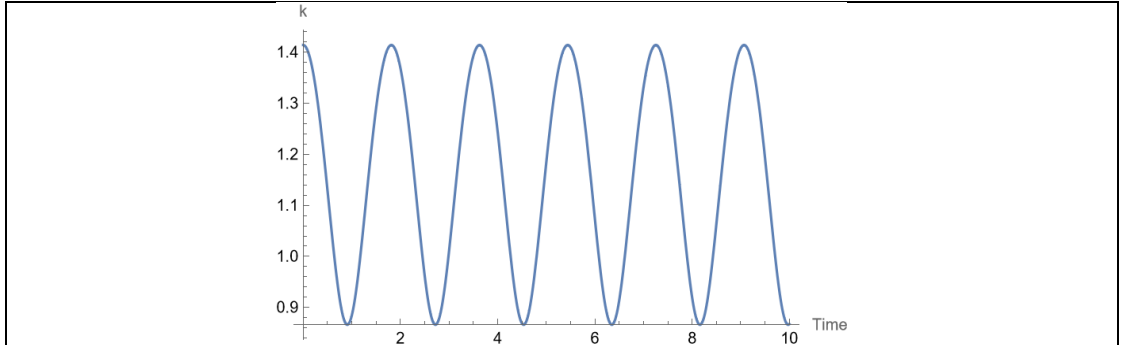

**Figure 3:** The localization parameter $k$ in the time interval $t \in [0,10]$, in the case that $\omega_1 = \sqrt{3}$, $\omega_2 = \sqrt{5}$, $\varphi_0 = 0$ and $\theta_0 = \pi/2$.

According to Eq. (14), the electromagnetic field corresponding to the parameter $k$ given by Eq. (37) is

$$\mathbf{E} = \frac{\omega_1 \omega_2}{2q}\left[\cos(\varphi_0 + \omega_2 t)\mathbf{i} + \sin(\varphi_0 + \omega_2 t)\mathbf{j}\right] \quad (38)$$
$$\mathbf{B} = \mathbf{0}$$

Thus, the localization of the particles is fully determined by the electromagnetic field in their region. This becomes more evident considering that one of the functions $\theta(t)$, $\varphi(t)$ is constant, while the other one is arbitrary. For example, supposing that $\theta(t)$ is constant, $\theta(t) = \theta_0$, the parameter $k$ becomes

$$k = \frac{1}{2}\left|\sin\theta_0 \frac{d\varphi}{dt}\right| \quad (39)$$

implying that

$$\frac{d\varphi}{dt} = \frac{2k}{\sin\theta_0}, \quad \sin\theta_0 \neq 0 \quad (40)$$

Here, it has been assumed that $\sin\theta_0$ is positive, which is true, since $\theta_0$ represents the polar angle taking values in the interval $(0, \pi)$. In this case the electromagnetic field (14) takes the simple form

$$\mathbf{E} = -\frac{1}{q\sin\theta_0}\frac{dk}{dt}\mathbf{k}, \quad \sin\theta_0 \neq 0 \quad (41)$$
$$\mathbf{B} = \mathbf{0}$$

On the other hand, assuming that $\varphi(t) = \varphi_0$ is constant, the parameter $k$ becomes

$$k = \frac{1}{2}\left|\frac{d\theta}{dt}\right| \quad (42)$$

and the electromagnetic field (14) takes the form

$$\mathbf{E} = \frac{1}{q}\frac{dk}{dt}\left(\sin\varphi_0 \mathbf{i} - \cos\varphi_0 \mathbf{j}\right)$$
$$\mathbf{B} = \mathbf{0}$$
(43)

In all cases, the parameters $\theta_0$, $\varphi_0$ represent o the polar and azimuthal angle of the particles at the time of the application of the fields. From the above expressions, it is clear that the rate of change of the localization of the particles can be fully determined by the electric field in their region and can take a constant value in regions with constant electric field. Also, the sign of the rate of change of the localization parameter $k$ depends on the direction of the field. Specifically, if the field is parallel (antiparallel) to the vector of the angular velocity of the particles, the parameter $k$ decreases (increases). The opposite is true for particles with negative helicity.

More generally, it can be easily verified that, if the particles move in a constant electric field $\mathbf{E}$ perpendicular to their direction of motion, the localization parameter $k$ changes at a constant rate equal to $|dk/dt| = q|\mathbf{E}|$. In S.I. units, the rate of change of $k$ becomes $|dk/dt| = q|\mathbf{E}|c$, where $c$ is the speed of light in vacuum. Assuming that the charge of the particles is equal to that of the electron, the above analysis implies that the localization parameter $k$ changes at a rate of $c|\mathbf{E}|$ $eV/s$, or $|\mathbf{E}|$ $eV$ per meter of propagation inside the constant field.

Thus, applying an electric field parallel to the propagation direction of Weyl particles, it is possible to change their energy, as discussed in the paragraph below Eq. (28), while applying a field perpendicular to their propagation direction, it is possible to change their localization, as discussed here. Consequently, the state of Weyl particles, regarding both their energy and localization can be manipulated using simple electric fields. Obviously, this is a particularly interesting result with important possible applications in materials supporting Weyl particles [8-18]. Here, it should be mentioned that the dynamics of Weyl particles can also be simulated using trapped ions by laser beams [19-24], providing an additional opportunity to study experimentally the predictions of our work.

In addition, it should be mentioned that the state of the particles, and consequently their localization, will not be affected if the electromagnetic fields given by Eq. (20) are added to the ones given by Eqs. (38), (41) and (43). Thus, there is a whole family of electromagnetic fields which could be utilized to control the localization of Weyl particles, as desired.

As an example, we consider a Weyl particle, where the initial values of the parameters $\omega_{01}$, $\omega_{02}$, $\varphi_0$ and $\theta_0$ are 0, 10, 0, $\pi/2$ respectively, corresponding to $k=5$. Thus, initially, the particle is localized, and the corresponding classical particle moves on a circle with radius $\omega_{02}^{-1} = 1/10$. Assuming that a constant electric field of the form $\mathbf{E} = (1/q)\mathbf{k}$ is applied, the rate of change of the parameter $k$ becomes $dk/dt = -1/2$. Consequently, the localization of the particle decreases and after a time interval of $t = 10$ it will become momentarily a free particle $(k = 0)$. Afterwards, the particle will become localized again, and at $t = 20$ the localization will retrieve its original value, as shown in figure 4.

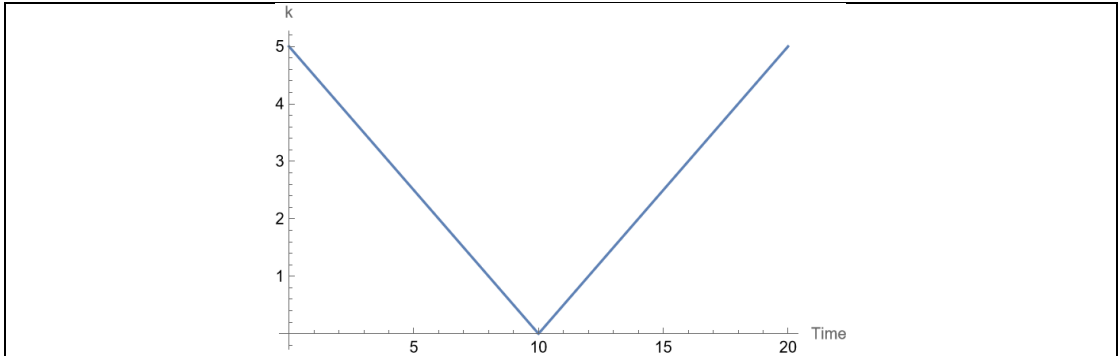

**Figure 4**: Evolution of the localization parameter $k$, for a Weyl particle with initial parameters $\omega_{01} = 0$, $\omega_{02} = 10$, $\varphi_0 = 0$ and $\theta_0 = \pi/2$, in a region of space with constant electric field $\mathbf{E} = (1/q)\mathbf{k}$.

The motion of a classical particle with velocity equal to the Weyl one is shown in figure 5. The delocalization and re-localization of the particle is evident.

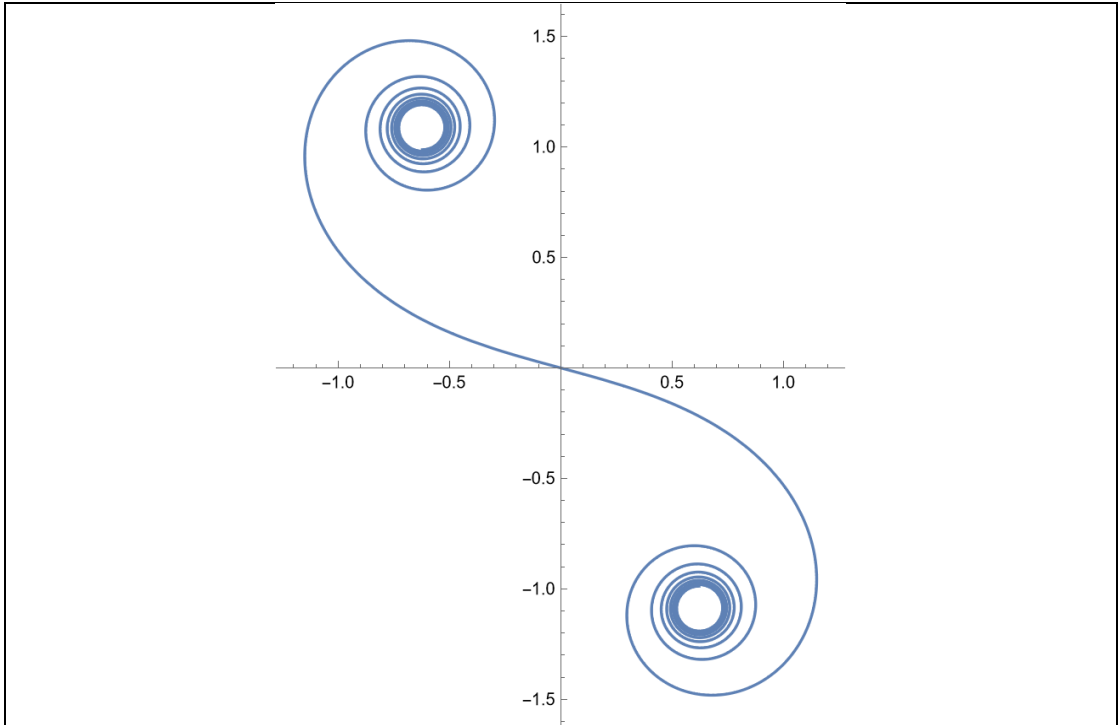

**Figure 5**: The motion of a classical particle with the same velocity as that of a Weyl particle with initial parameters $\omega_{01} = 0$, $\omega_{02} = 10$, $\varphi_0 = 0$ and $\theta_0 = \pi/2$, moving in a region of space with constant electric field $\mathbf{E} = (1/q)\mathbf{k}$.

## 4. On the connection between localization and mass

In this section we shall discuss the connection between the localization of the particles and the generation of gravitational mass. As discussed in the previous section, in the paragraph below Eq. (28), it is possible to control the energy of the particles through appropriate electric fields. At the same time, it is also possible to control the localization of the particles, through another family of fields, as detailed in the text below Eq. (43). Thus, the energy density produced by the localization of Weyl particles, can be used as a source term in the energy momentum tensor $T_{\mu\nu}$ in Einstein's field equations of general relativity, which, in S.I. units, can be written as [25-27]

$$G_{\mu\nu} \equiv R_{\mu\nu} - \frac{1}{2} R g_{\mu\nu} = \frac{8\pi G}{c^4} T_{\mu\nu} \qquad (44)$$

leading to the appearance of gravitational mass. Here, $G_{\mu\nu}$ is the Einstein tensor, which is a symmetric and divergence-free combination of the Ricci tensor $R_{\mu\nu}$ and the metric $g_{\mu\nu}$, and $R = g^{\mu\nu} R_{\mu\nu}$. It should also be mentioned that the Ricci tensor is related to the more general Riemann curvature tensor [28, 29] through the formula $R_{\mu\nu} = R^{\alpha}{}_{\mu\alpha\nu}$. In the above expressions all tensors are written in abstract index notation [27]. Also, the proportionality constant $8\pi G/c^4$ is used to assure that the weak-gravity, low-speed limit of general relativity is Newtonian mechanics. Here, $G$ is the gravitational constant.

Consequently, it becomes clear that the localization of Weyl particles leads to the appearance of a gravitational mass, equal to $E/c^2$ in S.I. units, where $E$ is the energy of the particles, which is a free parameter in our solutions. Thus, spinors (1), (6) can describe particles with arbitrary mass and localization. Therefore, Weyl particles could be considered as fundamental building blocks of the Universe, progenitors of a wide variety of particles.

It is also worth mentioning that, if the energy and localization of Weyl particles exceeds a critical value, they could even create a charged and rotating black hole, described by the Kerr–Newman metric [30-36]. For example, if the mass of the particle is of the order of the Planck mass, $m_P = \sqrt{\hbar c/G} = 2.176 \times 10^{-8} Kg$, corresponding to an energy of $E_P = m_P c^2 = 1.224 \times 10^{28} eV$, and it is localized in a region with dimensions of the order of the Planck length, $l_P = \sqrt{\hbar G/c^3} = 1.616 \times 10^{-35} m$, then it could create a tiny black hole. Furthermore, due to their small mass, the decay of these tiny black holes through the emission of Hawking radiation [37] is not allowed by the laws of quantum mechanics. Thus, these tiny black holes, the Planck relics as mentioned in the literature [38], are expected to be stable, offering a possible explanation for dark matter. Finally, it should be mentioned that, since Weyl quasi-particles have recently been observed in certain materials [8-18], one could manipulate the energy and localization of these particles using the electromagnetic fields described in the previous sections, in order to create and study these tiny black holes in the lab. As mentioned in section 3, these phenomena could also be simulated using ions trapped by laser beams [19-24].

## 5. Conclusions

In this work we have shown that Weyl particles can exist at different states in zero electromagnetic field, either as free particles or at localized states described by a parameter $k$ with dimensions of mass. In addition, it is possible to modify the localization of the particles, using simple electromagnetic fields, which are explicitly calculated. Consequently, the localization of Weyl particles in certain materials supporting these particles could be studied experimentally quite easily. It could also be simulated using ions trapped by laser beams. In addition, a particularly important remark is that the localization of the energy of Weyl particles can lead to the generation of real gravitational mass, through Einstein's field equations of general relativity. Thus, Weyl particles can be considered as fundamental building blocks of the universe, progenitors of a wide variety of particles. Furthermore, if their energy and localization exceed a critical level, they could even create tiny black holes, potential candidates for the black matter of the universe.